\documentclass[conference, 10pt]{IEEEtran}
\usepackage{amsmath,amssymb,amsfonts}
\usepackage{algorithmic}
\usepackage{graphicx}
\usepackage{tabularx}
\usepackage[normalem]{ulem}
\usepackage{textcomp}
\usepackage{xcolor}
\usepackage{listings}
\usepackage{subcaption}
\usepackage{biblatex}
\def\BibTeX{{\rm B\kern-.05em{\sc i\kern-.025em b}\kern-.08em
    T\kern-.1667em\lower.7ex\hbox{E}\kern-.125emX}}

\newcommand\jsonkey{\color{purple}}
\newcommand\jsonvalue{\color{cyan}}
\newcommand\jsonnumber{\color{orange}}

\makeatletter
\newif\ifisvalue@json

\lstdefinelanguage{json}{
    tabsize             = 4,
    showstringspaces    = false,
    keywords            = {false,true},
    alsoletter          = 0123456789.,
    morestring          = [s]{"}{"},
    stringstyle         = \jsonkey\ifisvalue@json\jsonvalue\fi,
    MoreSelectCharTable = \lst@DefSaveDef{`:}\colon@json{\enterMode@json},
    MoreSelectCharTable = \lst@DefSaveDef{`,}\comma@json{\exitMode@json{\comma@json}},
    MoreSelectCharTable = \lst@DefSaveDef{`\{}\bracket@json{\exitMode@json{\bracket@json}},
    basicstyle          = \ttfamily
}

\newcommand\enterMode@json{%
    \colon@json%
    \ifnum\lst@mode=\lst@Pmode%
        \global\isvalue@jsontrue%
    \fi
}

\newcommand\exitMode@json[1]{#1\global\isvalue@jsonfalse}

\lst@AddToHook{Output}{%
    \ifisvalue@json%
        \ifnum\lst@mode=\lst@Pmode%
            \def\lst@thestyle{\jsonnumber}%
        \fi
    \fi
    \lsthk@DetectKeywords%
}

\usepackage{graphicx}
\usepackage{etoolbox}
\newcommand{\insertfig}{\includegraphics[width=\linewidth]{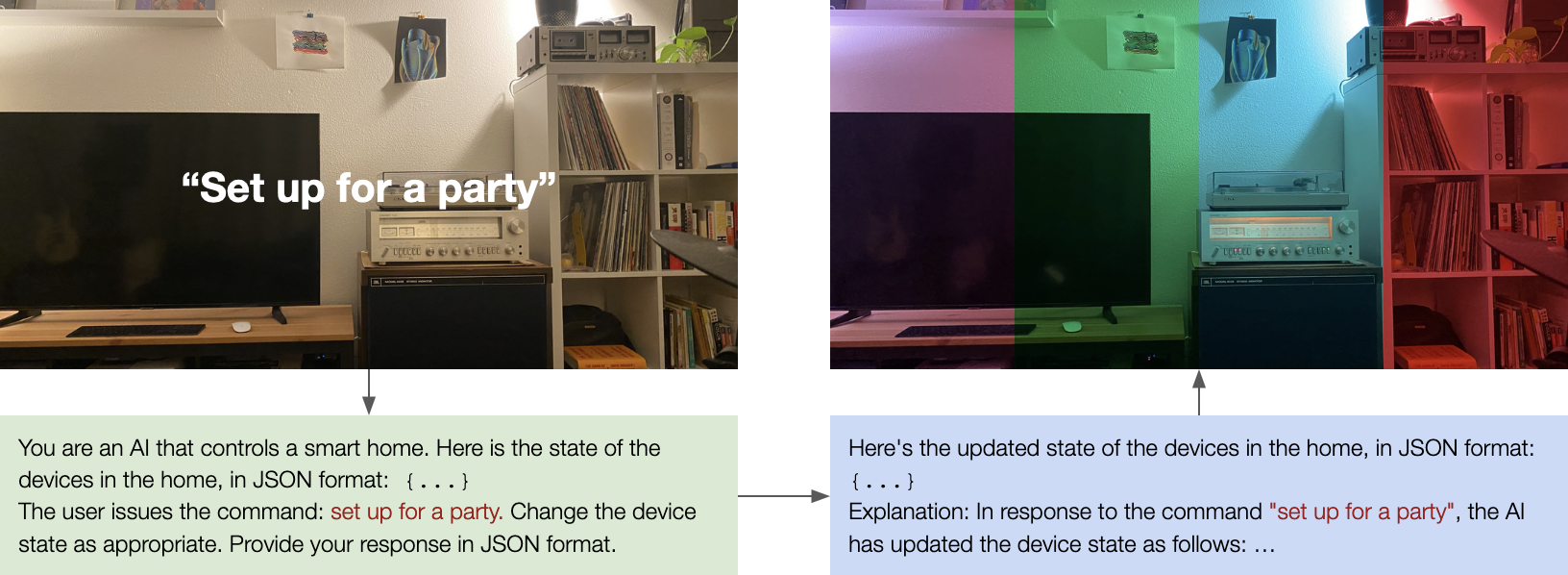}\captionof*{figure}{GPT models can infer meaning behind ambiguous user commands and control smart home devices in response. When told to ``set up for a party'', GPT-3 produces valid JSON that sets the lights to a color loop and turns on the stereo for music playback.}}

\makeatletter
\apptocmd{\@maketitle}{\centering\insertfig}{}{}
\makeatother

\begin{document}

\title{``Get ready for a party'': Exploring smarter smart spaces with help from large language models}

\author{
    \IEEEauthorblockN{Evan King, Haoxiang Yu, Sangsu Lee, Christine Julien}
    \IEEEauthorblockA{
        The University of Texas at Austin \\
        \{e.king, hxyu, sethlee, c.julien\}@utexas.edu
    }
}

\maketitle

\begin{abstract}
The right response to someone who says ``get ready for a party'' is deeply influenced by meaning and context. For a smart home assistant (e.g., Google Home), the ideal response might be to survey the available devices in the home and change their state to create a festive atmosphere. Current practical systems cannot service such requests since they require the ability to (1)~infer meaning behind an abstract statement and (2)~map that inference to a concrete course of action appropriate for the context (e.g., changing the settings of specific devices). In this paper, we leverage the observation that recent task-agnostic large language models (LLMs) like GPT-3 embody a vast amount of cross-domain, sometimes unpredictable contextual knowledge that existing rule-based home assistant systems lack, which can make them powerful tools for inferring user intent and generating appropriate context-dependent responses during smart home interactions. We first explore the feasibility of a system that places an LLM at the center of command inference and action planning, showing that LLMs have the capacity to infer intent behind vague, context-dependent commands like ``get ready for a party'' and respond with concrete, machine-parseable instructions that can be used to control smart devices. We furthermore demonstrate a proof-of-concept implementation that puts an LLM in control of real devices, showing its ability to infer intent and change device state appropriately \emph{with no fine-tuning or task-specific training}. Our work hints at the promise of LLM-driven systems for context-awareness in smart environments, motivating future research in this area.
\end{abstract}


\section{Introduction}
\label{introduction}
An exciting prospect of smart homes at their advent was the potential to reduce user burden by providing seamless, unobtrusive, and ``smart'' interfaces to everyday devices. While smart assistants have improved significantly over the years with respect to speech recognition~\cite{swarup2019improving, raju2019scalable} and user satisfaction~\cite{purington2017alexa, lopatovska2019talk}, a central challenge remains: how can these assistants be made to respond appropriately to ambiguous user commands that may be influenced by context or are otherwise impossible for system developers to anticipate beforehand? An example of such a command might be a user preparing their home to entertain for guests, who asks their smart assistant to ``get ready for a party''. The hope is that the assistant---if it is truly smart---might be able to help by inferring the meaning of the statement and determining how to change the state of available devices in response: perhaps to start up the user's party playlist on a smart speaker and change their smart lights to a festive color scheme. In practice, however, such a request is beyond the capacity of current smart home systems. Google Home will sadly admit: ``I'm sorry, I didn't understand.''

In this paper, we are motivated by the observation that large language models (LLMs) like OpenAI's GPT-3~\cite{brown2020language} have shown an impressive ability to generalize to new tasks with high zero-shot performance, as well as the capacity to infer meaning behind semantically complex or abstract statements~\cite{liu2021gpt}. We thus ask the question: can this powerful capacity for cross-domain contextual reasoning be applied to practical smart home applications? 

To explore this question, we carry out a feasibility study that places GPT-3 in control of a smart home. We evaluate GPT-3's ability to provide high-quality responses to user commands of varying ambiguity given only a simple prompt and a data structure containing information about devices that it can control. Our results demonstrate that LLMs like GPT can infer the meaning behind ambiguous user commands like ``get ready for a party'' or ``I am tired and I want to sleep'' and respond with properly-formatted data describing courses of action, enabling more intuitive control of smart devices. We furthermore build a proof-of-concept implementation that puts GPT-3 in control of real devices, showing LLM-driven command inference and action planning can function in practice \emph{with no fine-tuning or task-specific training required}. Motivated by our results, we propose future work that can further leverage the power of LLMs toward building smarter smart home applications.

Our key contributions are as follows:

\begin{itemize}
    \item An experimental setup and study results that show LLMs can infer meaning behind abstract user commands like ``I am tired and I have to work'' and, in response, quickly and appropriately change the state of the smart devices available in the home, \emph{with no task-specific training}.
    \item An implementation that puts a GPT model in control of real devices, showing that it can intuitively respond to a variety of commands. When told to ``set up for a party'', it responds by turning on a stereo and configuring a group of Hue lights to loop through a festive set of colors; given the command ``I'm leaving'', it turns off all available devices. We trigger these actions by inputting the LLM's response \emph{directly into smart device APIs}.
    \item Analytical results that suggest responses are variable in quality, dependent on both the devices available and the nature of the user's command. In essence, further system design is necessary to manage the LLM's tendency to ``not know what it doesn't know'' in order to produce consistently high-quality responses.
\end{itemize}

The following describes the structure of this paper. Section~\ref{related-work} situates our work with related research. Section~\ref{approach} describes the experimental setup that we use to demonstrate the feasibility of LLMs as smart home controllers. Section~\ref{evaluation} presents the results of our exploratory study, while Section~\ref{implementation} demonstrates a proof-of-concept implementation. Section~\ref{limitations} offers avenues for future work. Section~\ref{conclusion} concludes.

\section{Background \& Related Work}
\label{related-work}
This section provides a high-level overview of LLMs and their applications before situating our work with related efforts in context-aware smart spaces.

\textbf{Large language models (LLMs)} have gained significant attention in recent years due to their impressive performance on a wide range of natural language processing tasks. In 2018, Devlin proposed BERT, a language representation model that uses Bidirectional Encoder Representations from Transformers~\cite{devlin2018bert} and can be fine-tuned for a variety of NLP tasks, such as text classification and sentiment analysis. In the same year, OpenAI proposed GPT (Generative Pre-trained Transformer)~\cite{radford2018improving}. Both models use a transformer architecture~\cite{vaswani2017attention} that was pre-trained on a massive corpus of text data, including books, articles, and websites. The resulting models demonstrate impressive results on a wide range of natural language processing tasks, including language translation, text generation, and the ability to translate natural language descriptions into program implementations.
 
Following the success of the transformer-based model, subsequent studies have explored ways to improve and expand the model's performance. In 2019, Radford et al., published an updated version of GPT and called GPT-2 \cite{radford2019language}. Building on the success of GPT-2, Brown et al. released GPT-3 in 2020 \cite{gpt3}. After that, in 2023, GPT-4 was introduced. It is currently one of the largest and most powerful language models, with more than 1 trillion parameters \cite{gpt_4}. At time of writing, access to GPT-4 is limited---we therefore base our study on GPT-3.

Two popular approaches exist for adapting task-agnostic LLMs to new applications: \emph{prompt engineering} and \emph{fine-tuning}. Prompt engineering refers to the process of designing a task-specific prompt or template that guides the model to produce relevant outputs for a particular task~\cite{zhao2021calibrate}. These prompts generally contain instructions to the model written in natural language---e.g., ``explain the following passage of text''. Fine-tuning, on the other hand, involves directly training the model on a new task by providing task-specific examples~\cite{radford2019language}. The key advantage of prompt engineering over fine-tuning is that it does not require task-specific data---we therefore adopt that approach here. Within the realm of prompt engineering, there are \emph{zero-shot} and \emph{few-shot} learning approaches. Zero-shot approaches provide the model with a single prompt containing instructions and task-specific information; few-shot approaches provide examples to the model of correct input/output pairs. We focus on zero-shot learning.

\textbf{Context-aware spaces} leverage sensor information, user data (including past behaviors and preferences), and device state to influence system actions toward meeting user needs~\cite{6177682}. The notion of ``context-awareness'' in this sense has roots in research on ambient intelligence~\cite{cook2009ambient}---that is, the development of built environments that sense and adapt to users. A concrete example of this concept is a home that leverages contextual information to improve energy efficiency~\cite{jahn2010energy, geraldo2019energy}. In an early paper, Yamazaki suggested that smart homes should go beyond automation and instead integrate expressive interfaces between the user and system~\cite{yamazaki06}, a goal that is partially realized in smart assistants~\cite{purington2017alexa}, but with limited ability to adapt to more complex user commands~\cite{kiseleva2016understanding}. Ample prior work has approached the issue of context-awareness using task-specific models~\cite{qolomany2019leveraging, kabir2015machine, machorro2020hems, liang2018unsupervised}. While these methods can achieve high performance given ample task-specific data, we believe that the high zero-shot performance of LLMs could hint at better generalizability without a need for training data. However, we are aware of no work to-date that has explored the use of task-agnostic LLMs for deeper contextual reasoning in smart environments. This motivates our feasibility study.

\section{System Design}
\label{approach}
In this section, we introduce the system design that we use to explore the feasibility of LLM-driven smart home control. We first assume the use of an LLM like GPT-3 that provides responses to user prompts written in natural language. These LLM models are not task-specific, rather, they are trained on an immense amount of cross-domain textual information and, depending on the structure of the prompt, can provide responses suited to a variety of different use cases (e.g., writing a poem, writing code in response to a high-level program description, etc.). We opt to adapt the model's outputs to our task using zero-shot learning through prompt engineering.

Our challenge is therefore to package relevant context 
and user commands into a concise prompt issued to the model, such that its responses include concrete, machine-parseable changes to device state that can be passed off to the appropriate smart device APIs. Qualitatively, we want these courses of action to be shaped by the model successfully inferring (1)~the intent behind the user command and (2)~the manner in which the state of available devices can be changed to meet the user's intent. To that end, we first define an abstract schema for capturing smart home context before describing a method for engineering prompts to conversational LLMs that elicit useful, actionable responses.

\subsection{Context Schema}
In order for the model to ``know'' what actions are available to it, we need to package the available devices, their states, and other relevant information---i.e., the context---into a machine-parseable format. This package effectively describes the action space available to the model: the knobs it can turn, and information (e.g., which room the user is in) that might influence {\it how} it turns them. It also provides a hint about how the model should format its response. Representations of context can be complex and have been explored in the literature \cite{hua2022copi, warble21}. Since our goal is to conduct an exploratory study rather than design an end-to-end solution, we use a schema that is simple but adequate for our experimental setup. We choose JSON for structuring this data since it is the de-facto data interchange format for RESTful APIs used by many smart home devices~\cite{hue2022, nest2022, insteon2022}. Leveraging a common format is also advantageous since there is a high likelihood that the LLM has been trained on source material that contains it, which benefits the model's ability to converse in it.

\begin{figure}
    \centering
    \includegraphics[width=\columnwidth]{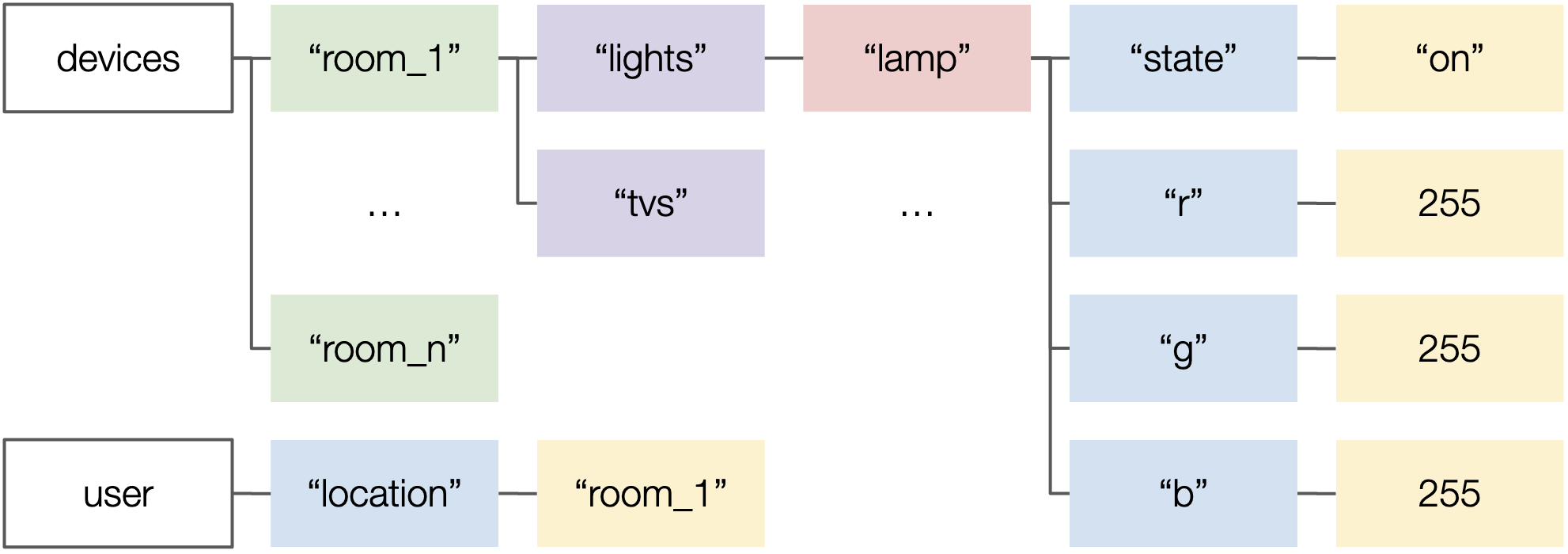}
    \caption{Data structures for expressing smart home device and user context in prompts to an LLM.}
    \label{fig:schema}
\end{figure}

At the top level, context is a collection of ``key, value'' pairs. There are two relevant contexts: ``user'' context that contains immutable information about the user's state, e.g., which room they are in, and ``device'' context, which contains mutable and immutable information about the devices in the home. Each top-level key in the collection of device context defines a room in the home, and within each room we define collections of devices organized by type, e.g., ``lights'', ``tvs'', etc. Within a collection of devices, we define each individual device as a collection of properties about that device, e.g., for a light we can define its ``state'' property and its ``r'', ``g'', and ``b'' color values. This overall structure is depicted in Fig.~\ref{fig:schema} and illustrated by the example in the following:

\begin{lstlisting}[language=json]
{
  "user": {
    "location": "living_room"
  }
}
\end{lstlisting}
\vspace{-.2cm}
\begin{lstlisting}[language=json]
{
  "devices": {
    "bedroom": {
      "lights": {
        "bedside_lamp": {
          "state": "off"
        }
      }
    },
    "living_room": {
      "lights": {
        "overhead": {
          "state": "on"
        },
        "lamp": {
          "state": "off"
        }
      },
      "tvs": {
        "living_room_tv": {
          "state": "off",
          "volume": 20
        }
      }
    }
  }
}
\end{lstlisting}

In this example, the user's home has two rooms---a bedroom and living room---and the user is currently located in the living room. The bedroom has one light turned off, and the living room has two lights (one turned on) and a television.

\subsection{Prompt Engineering}
Having developed a structure for storing context, we now move to the practical challenge of engineering  prompts that elicit useful responses from the model.

Our prompts consist of four segments, as follows:
\begin{itemize}
    \item \textbf{Framing.} This portion of the prompt provides direction to the conversational agent about its role in the interaction---it is being asked to make decisions as an AI that controls a smart home. We open with the phrase ``You are an AI that controls a smart home.''
    \item \textbf{Context.} This informs the agent of the user context and devices available in the environment, which scopes the space of its actions and provides a hint as to the structure of our desired response. We continue the prompt: ``Here is the state of the devices in the home, in JSON format: \{devices\} Here is information about the user: \{user\}'', where both contexts are formatted as shown earlier.
    \item \textbf{Command.} This portion inserts the user command and directs the agent to manipulate the state of devices in response, as follows: ``The user issues the command: \{command\}. Change the device state as appropriate.'' The command is written in natural language, as a user might utter to their smart assistant.
    \item \textbf{Formatting.} We close the prompt by requesting the response in JSON format so that it can be easily parsed and input to a relevant smart device API: ``Provide your response in JSON format.''
\end{itemize}

\begin{figure}
    \centering
    \includegraphics[width=\columnwidth]{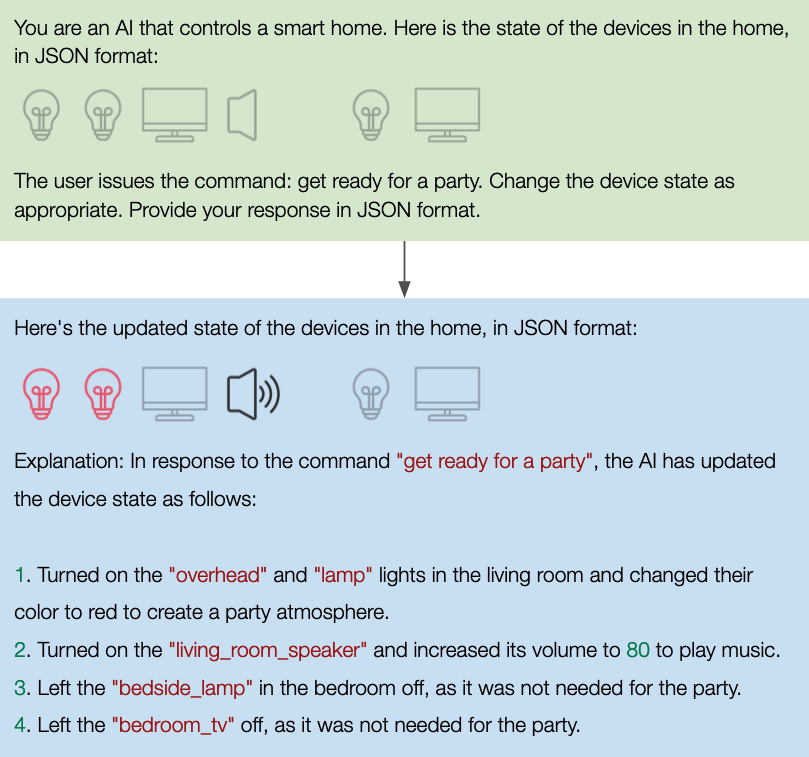}
    \caption{An example prompt and response from ChatGPT, demonstrating its ability to change device state in response to ambiguous user commands like ``get ready for a party''. JSON is omitted from this figure in favor of a visual depiction. }
    \label{fig:prompt_example}
\end{figure}

An example prompt with this structure and the corresponding response from ChatGPT 3.5 are depicted in Figure~\ref{fig:prompt_example}. We can see that by using the proposed context structure inside the the prompt, we are able to elicit responses from the model that contain changes to the underlying JSON that accurately reflect what a user's intent might be. In essence, GPT-3.5 is able to relate the meaning of ``party'' to the devices available, as well as alter their specific settings in desirable ways. In the next section, we use this system design to perform qualitative analysis of the model's responses.

\section{Evaluation}
\label{evaluation}
This section describes the results of our feasibility study using the experimental setup described in the previous section. Our evaluations address two high-level questions:

\begin{enumerate}
    \item \textbf{How good are the agent's responses?} We measure the quality of the agent's responses, in the sense that they include courses of action that can reasonably be thought to meet the user's request and can be easily machine-parsed and executed.
    \item \textbf{How timely are the agent's responses?} We also measure the round-trip response latency. This hints at how feasible a practical system is with respect to user experience and responsiveness.
\end{enumerate}

To better understand the system from these two perspectives, we design scenarios of increasing complexity and ambiguity of \emph{context} and \emph{command}. This captures the intuition that (1)~different smart homes can have different complexity of context, from an apartment with a few smart lights to a large home with many devices and (2)~different user commands can have different levels of ambiguity, from direct commands like ``turn on the light'' to wholly ambiguous statements like ``I am tired''. Evaluating agent responses under these circumstances allows us to identify the failure modes of LLM-driven smart home control given increasingly challenging prompts with respect to both the context and the nature of the command. 

We use three contexts of increasing complexity, as follows:

\begin{itemize}
    \item \textbf{Simple:} Describes a home with a bedroom and living room that have one and two lights, respectively, all initially off. Lights can either be on or off but have no other state (e.g. color).
    \item \textbf{Medium:} Same as above, but adds red, green, and blue color state to each of the lights, with expected values in the range [0, 255].
    \item \textbf{Complex:} Same as above, but adds a TV with on/off and volume state to the bedroom, as well as a TV and smart speaker to the living room (each also with on/off and volume state).
\end{itemize}

Each of these contexts is expressed in the schema described in Section~\ref{approach}. We combine these contexts with three user prompts of increasing ambiguity, as follows:

\begin{itemize}
    \item \textbf{Direct:} {\it ``Turn on the light.''} This command is simple since it directly expresses a state change, as well as a relevant device. Existing home assistants can easily respond to this type of command.
    \item \textbf{Indirect:} {\it ``Get ready for a party.''} This command is more ambiguous since it expresses a desired state change, but provides no information about which devices are relevant.
    \item \textbf{Ambiguous:} {\it ``I am tired.''} This command is completely ambiguous since it expresses neither a state change, nor which devices might be relevant.
\end{itemize}

We run our tests with each possible combination of these three contexts and commands (9 total), each for 10 trials. We save the agent's response for each trial in a human-readable format, then perform manual rating to measure the quality of the responses. Our process for rating the quality of responses is based on a binary label, where each is assigned one of the following labels based on its quality:

\begin{itemize}
    \item \textbf{Poor (0):} ``The changes to the devices do not at all reflect the intent behind the user command, or the response is malformed/garbled.''
    \item \textbf{Good (1):} ``The changes to the devices are reasonable for the command. You can imagine \emph{someone} being satisfied with the result, even if it is somewhat subjective (e.g., based on different personal preferences).''
\end{itemize}

Three researchers independently reviewed all responses and assigned them a label. We report the aggregate score for each trial as the average across all assigned scores. We also note the average latency for each trial---this includes both the network transmission time of the request, as well as the inference time taken by the model. Since this time is subject to network conditions and API demand, it should be taken as a rough estimate rather than a concrete benchmark. Our results are summarized in Table~\ref{tab:results}.

\textbf{Response time is a function of context complexity.} With respect to latency, we can see that responses generally arrive on the order of seconds, meaning that a practical system could feasibly leverage an LLM for ambiguous command inference and action planning without significant detriments to user experience or responsiveness. For direct commands, a 2 to 3 second response time may be too long---future system designs could thus leverage the LLM only for commands that require it. This may entail a hybrid of rule-based inference for common commands, along with LLM inference for less familiar commands. It is also worth noting that as the context increases in complexity, the response latency also increases---this motivates future work to develop methods for filtering context prior to prompting the agent so that only the most relevant information is provided. 

\textbf{Response quality is a function of context and command ambiguity.} With respect to response quality, we find that the LLM approach provides good responses given the same direct and simple commands that current home assistants are able to service. Note, however, that unlike existing home assistants, the LLM approach utilizes a much simpler system architecture that performs command inference and action planning in the same pass. These results are consistent given increasing degrees of context complexity, suggesting that the model was not overwhelmed by the growth in the decision space that comes with adding new devices and possible state changes. On the contrary, the model provides \emph{better} responses when given more context that might be relevant to the user's command, as is apparent when comparing the low response quality of the Simple/Indirect and Medium/Indirect experiments against the Complex/Indirect experiment. Upon inspection of the responses, the reason for this is clear: given minimal context and a subjective command like ``get ready for a party'', the LLM simply makes up a response---specifically, it turns on all the lights in the house, to include the bedroom. When we add a speaker and a television to the context, the model now has more relevant knobs to turn, and produces a higher quality response. The tendency to make something up when the answer is unknown or requires more context is an open problem and motivates the development of application-specific methods to mediate between the user and the model. 

For the most ambiguous command (``I am tired''), we note that the model delivers poor responses regardless of context. In all but a few cases, the LLM simply turns on all of the home's lights. The exception is in a Medium/Ambiguous trial, where it only turns on the bedroom lamp, perhaps to help the user prepare for bed. This is to be expected: an individual's intent and preference in this case are highly subjective (are they, e.g., tired and ready for bed or tired but they have a pressing deadline?) and the LLM ultimately cannot read the user's mind. However, since the LLM does not ``know what it doesn't know'', it does not ask for clarification or go with the safest choice, which is likely to do nothing. Instead, it makes something up.

Since our previous results suggested more context is often beneficial, we dig deeper to see if we can help it make a better choice. We amend the vague command ``I am tired'' to offer hints at the user's intent:

\begin{itemize}
    \item \textbf{Ambiguous*:} ``I am tired and I need to work.''
    \item \textbf{Ambiguous**:} ``I am tired and I want to sleep.''
\end{itemize}

Provided this added hint at the user's context, the response quality improves significantly. For Ambiguous*, the model consistently responds by turning on the living room lights while leaving all other devices off; for Ambiguous**, the model turns on only the user's bedside lamp and, in some cases, reduces the volume on their speaker and TV. Note that although the amended statement includes additional context, it still requires the model to infer meaning in a way that more rigid or rule-based approaches cannot.

\begin{table}[t!]
  \centering
  \begin{tabular}{|l | l | l | l|} \hline
    \textbf{Context} & \textbf{Command} & \textbf{Avg. Quality} & \textbf{Avg Latency (sec)} \\ \hline
    Simple  &  Direct & 1.00  & 2.42 \\
      &  Indirect &  0.67 & 2.31 \\
      &  Ambiguous & 0.00 & 2.22 \\ \hline \hline
    Medium  &  Direct & 1.00 & 4.56 \\
      &  Indirect & 0.63 & 4.70 \\
      &  Ambiguous & 0.17 & 4.97 \\ \hline \hline
    Complex  &  Direct & 1.00 & 7.90 \\
      &  Indirect & 1.00 & 7.25 \\
      &  Ambiguous & 0.00 & 7.04 \\ \hline \hline \hline
    Complex & Ambiguous* & 1.00 & 7.49 \\
     & Ambiguous** & 1.00 & 8.09 \\
    \hline
  \end{tabular}
  \caption{Results for experiments given various combinations of different context complexity and command ambiguity. Higher quality responses suggest the model produced a course of action that would be desirable for an end user (e.g., turning on the bedroom light when receiving the command ``I am tired and I want to sleep''). Lower latency suggests better system responsiveness.}
  \label{tab:results}
\end{table}

\section{Implementation}
\label{implementation}
To demonstrate LLM-driven smart home control in practice, we built a proof-of-concept implementation in Python. Our implementation accepts user commands as strings, packages them into prompts along with contextual information about real devices, then processes responses from OpenAI's \texttt{text-davinci-003} model into smart home API calls that change the device state as specified by the LLM. We scope the application to one room (a researcher's living room) with three Philips Hue color smart lights~\cite{hue2022} and one TP-Link Kasa smart plug~\cite{kasa2023} that controls a stereo. We store the device context in JSON as in our experimental setup, with the difference that the device state for the Hue lights is pulled directly from the Hue API without modification. Our code is open source and available online.\footnote{https://github.com/UT-MPC/homegpt}

The teaser figure depicts the result when issuing the command ``set up for a party''. We include the JSON context of the light group\footnote{https://developers.meethue.com/develop/hue-api/groupds-api/}, along with a field for the plug powering the stereo. The model mutates the parameters in the JSON to change the stereo state to ``on'' and, impressively, also changes the ``effect'' parameter of the Hue light group from ``none'' to ``colorloop'' to create a looping color effect. The latter change suggests that GPT-3 may have been trained on material about the specific features of the Hue API and can leverage that along with the inferred meaning of the user command to trigger more intuitive changes than existing systems.

We briefly list multiple other commands we tested in our implementation, along with responses from the model:

\begin{itemize}
    \item ``make it bright in here'' -- sets lights to full brightness
    \item ``make it groovy'' -- sets lights to color loop; adds invalid ``genre'' field to stereo and sets it to ''groovy''
    \item ``gotta relax'' -- dims lights, turns stereo on
    \item ``I'm cold'' -- sets lights to warm white, turns stereo on
    \item ``I'm leaving'' -- turns off lights and stereo
    \item ``I'm home'' -- turns on lights and stereo
\end{itemize}

We note, of course, that these tests are far from exhaustive. We observed high variability in responses, meaning the same command can elicit many responses: some good, some bad. A more robust system design will be necessary to tackle the inconsistencies present in current LLM model outputs. We address this in our discussion in the following section.

\section{Discussion \& Future Work}
\label{limitations}
Our efforts in this paper hint at exciting opportunities for future work. We suggest several avenues for further research.

\textbf{Managing contextual information.} We found that including more context can improve the quality of the model's responses, but at the expense of response latency. To effectively navigate this tradeoff in an end-to-end solution, a more involved approach for storing, pre-processing, and expressing context will be necessary. This will also become essential as the amount of context grows to include sensor data, user preference data, and a growing and more diverse set of controllable devices. We note that in our experiments, we did not attempt to test the limits of \emph{how much} context a model can receive before the quality or latency of responses degrades substantially. This should be considered in future work.

\textbf{Robust system design.} While we were able to leverage a simple system design in this paper, an end-to-end system will need a more robust design to account for several factors. First, since LLMs do not yet ``know what they don't know'', the likelihood of invalid or low-quality responses remains high. In the case of responses where the model makes invalid changes to device state (e.g, to add new settings to a device), a full system should include a way to enforce a set of formal properties for device states. In the case of unsatisfactory responses, it would be beneficial to develop a method for learning user preferences or seeking clarifying information (e.g., ``are you tired and want to sleep, or are you tired but need an energy boost?''). 

\textbf{From commands to automation.} Our primary focus in this exploratory study was on immediate commands---the user makes a request and the model immediately responds with a state change. Future work could investigate the use of LLMs for more intuitive automation planning. A user could, for instance, ask their smart assistant to ``play jazz when it rains'' and the model could leverage contextual information to put in place an automation sequence that meets their needs. This would obviate the need for pre-programmed automation routines and could substantially improve user satisfaction with smart assistant systems.

\section{Conclusion}
\label{conclusion}
In this paper, we explored the feasibility of smarter smart home control using large language models (LLMs). We proposed a simple system design for capturing smart home context (i.e., information about the user and controllable devices in the environment) in engineered prompts to GPT-3, showing that the model has the ability to infer meaning behind indirect and ambiguous user commands like ``I am tired and I need to work'' and, in response, generate changes to smart device state. We implemented our system design, giving GPT-3 control of real devices and finding that it is able to quickly and appropriately control them in response to user commands with no fine tuning and no post-processing of its responses. By simply telling GPT-3 what devices are available and what the user wants, it can generate courses of action in response.

Our work hints at the capability of GPT-3 and similar models to go far beyond the current abilities of smart space control and motivates future work with context modeling, end-to-end system design, and approaches for further leveraging GPT-3's capabilites to develop complex automation routines in response to user commands.

\printbibliography

\end{document}